\documentstyle[12pt]{article}
\begin{document}
\centerline{
FERMIONIC ENTROPY of the VORTEX STATE in $d$-WAVE
  SUPERCONDUCTORS.}
\vskip 10mm
\centerline{
G.E. Volovik}
\vskip 10mm
\centerline{Low Temperature Laboratory,
Helsinki University of Technology,}
\centerline{Otakaari 3A, 02150 Espoo, Finland}
\centerline{and}
\centerline{L.D. Landau Institute for Theoretical Physics}
\centerline{
Kosygin Str. 2, 117940 Moscow, Russia}
\centerline{and}
\centerline{CRTBT, CNRS, B.P.166 25, Avenue des Martyrs,}
\centerline{ 38042 Grenoble CEDEX 09, France
}

\vskip 10mm
\begin{abstract}
{In the $d$-wave superconductors the electronic entropy
associated with an isolated vortex diverges logarithmically
with  the size of the system even at low temperature. In the
vortex array the entropy per vortex per layer, $S_V$, is
essentially larger than $k_B$ and depends on the
distribution of the velocity field $v_s$ around the vortex.
If there is a first order transition with the change of the
velocity distribution, then there will be a big entropy jump
$\Delta S_V \sim 1 k_B$ at the transition.
This entropy jump comes from the electronic degrees of
freedom on the vortex background, which is modified by the
vortex transition. This can explain the big jump in the
entropy observed in the so-called vortex-melting transition
\cite{Junod}, in which the vortex array and thus the
velocity field are redistributed. The possibility of the
Berezinskii-Kosterlitz-Thouless transition in the 3-
dimensional $d$-wave superconductor due to the fermionic
bound states in the vortex background is discussed.}
\end{abstract}

\vskip 5mm
Pis'ma ZhETF, {\bf 65},  465  (1997) [JETP Lett. {\bf
65},  491  (1997)]
\vfill\eject

\section{Gap nodes and scaling.}

The low energy properties of the superconductors with nodes
in the energy gap are governed by the electronic excitations
close to the gap nodes. The electronic   density of states
(DOS) in the homogeneous superconductor is (see eg the
Review paper \cite{Ankara})
$$N(E) \sim N_F \left({E\over T_c}\right)^{2-D}~~,~~ E\ll
T_c, \eqno(1)$$
where $D$ is the dimension of nodes and $N_F$ is the DOS on
the Fermi level in the normal metal. In the mixed state of
superconductor the superflow around the vortex leads to the
Doppler shift of the energy  $E$  to
$E + {\bf v}_s({\bf r})\cdot {\bf p}$, where
 ${\bf v}_s({\bf r})$ is the local superfluid velocity.
This gives the finite DOS at zero energy
$$N(0) \sim N_F \left({p_F v_s\over T_c}\right)^{2-D}
.\eqno(2)$$
Here $v_s$  is the characteristic value of the superfluid
velocity in the vortex array: $v_s\sim \hbar /m_3 R_V$,
where $R_V\sim \xi \left({B_{c2}\over B}\right)^{1/2}$ is
the intervortex distance, $B$ is magnetic field and $\xi
\sim v_F/T_c$ is the coherence length. Thus
$p_F v_s/  T_c \sim \sqrt{B\over B_{c2}}$ and this gives the
following electronic DOS  at zero energy as a function of
magnetic field
 $$N(0) \sim N_F \left({B\over B_{c2}}\right)^{1-D/2}
.\eqno(3)$$

There are two different regimes of strong and weak fields
with the crossover parameter
$$x\sim {T\over k_F v_s}\sim {T\over T_c}\left({B_{c2}\over
B}\right)^{1/2}.  \eqno(4)$$
which separates the superflow  dominating regime $x\ll 1$
from the temperature dominating regime $x\gg 1$ (see
\cite{KopninVolovik}).
In general the thermodynamic functions depend on the
parameter $x$. For example, the free energy of the
excitations on the background of the vortex array is
$$F(T,B) = N_F T_c^2 \left({B\over B_{c2}}\right)^{2-D/2}
\tilde F(x) \eqno(5)$$
where $\tilde F(x) $ is the dimensionless function of the
dimensionless parameter $x$ (see also the
recent paper by Simon and Lee \cite{SimonLee}; their crossover
parameter differs from our  by the factor
$\sqrt{T_c/E_F}$. This is because Simon and Lee used
the linearized spectrum of the fermions in the very vicinity
of the gap nodes,  which can be justified only at rather low
temperature, $T\ll T_c^2/E_F$). The normalization can be found
from the low field asymptote
$x\gg 1$, where the largest contribution comes from the bulk
superconductor, while the effect of vortices is small. It
follows from Eq.(1) that the free energy of the homogeneous
state is $\sim - N_F T_c^2 (T/T_{c2})^{4-D}$, and this gives
the normalization of $F(T,B)$ and the low field asymptote
$$\tilde F(x)\sim -x^{4-D} ~~,~~ x\gg 1. \eqno(6)$$

\section{Scaling for $d$-wave superconductor.}
\subsection{Free energy}

In the case of
nodal lines, ie for a node dimension $D=1$, one has
the  following estimate for the free energy of the
excitations  in the background of the vortex array:
$$F(T,B) = N_F T_c^2\left({B\over B_{c2}}\right)^{3/2}
\tilde F(x) \eqno(7)$$

Let us find the asymptotes of $\tilde F(x) $ at $x\gg 1$ and
$x\ll 1$.
Let us consider first the weak-field case  $x\gg 1$,
i.e., the case of a dilute vortex array. In addition to the
largest  asymptote
$\tilde F(x)\sim -x^3 $ from the bulk superconductor there
is also the contribution from vortices. It comes from the
modification of the normal-component density due to
excitations: $\rho_n(T) \sim \rho
N(T) / N_F$. This leads to a decrease of the kinetic energy of
superflow around the vortex, thus the contribution to the
energy of the vortex array with the vortex density
$n=B/\Phi_0$, where $\Phi_0$ is the flux quantum, is
$$F (B \rightarrow 0)=- {1\over 2}n\rho_n(T)\int_{R_V>r>
\hbar v_F/T}
d^2r~{\bf v}_s^2 \sim - N_F T_c^2 {B\over B_{c2}}{T\over
T_c}   \ln {R_V T\over \xi T_c}
~~.\eqno(8)$$
This corresponds to the term  $-x\ln x$ in $\tilde F(x)$.
So, the leading terms in the asymptote  of $\tilde F(x) $ at
$x\gg 1$ are
$$ \tilde F(x) \sim -x\ln x - x^3 ~~,~~x\gg 1.\eqno(9)$$

The asymptote of $\tilde F(x) $ at $x\ll 1$ is
$$ \tilde F(x) \sim -1 - x^2 ~~,~~x\ll 1\eqno(10)$$
Here both terms are from the vortices. The first
one is temperature independent and does not contribute
to the entropy or specific heat, but does contribute to the
vortex magnetization. It comes from the nonzero
normal-component  density at $T=0$ due to the superfluid
velocity:
$\rho_n(T=0) \sim \rho
N(E=0) / N_F \sim \rho (p_F v_s/ T_c)^{2-D}$ (see
\cite{VolovikMineev1981,Muzikar1983,Nagai1984,Xu1995}):
$$F(T \rightarrow 0) =- {1\over 2}n\rho\int_{R_V>r> \xi}
d^2r~{\bf v}_s^2 \left({p_F v_s\over  T_c}\right)^{2-D}\sim
-N_F T_c^2 \left({B\over B_{c2}}\right)^{3/2}
~~.\eqno(11)$$

The second (quadratic in $x$) term in Eq.(10) gives a term
linear in temperature to  the specific heat: $C(T,B)
\propto T \sqrt B$ \cite{d-waveVortex}.

\subsection{Vortex entropy}

For the entropy density one has
$$S(T,B) =-{\partial F\over \partial T}=
 N_F  {B\over B_{c2}} T_c \tilde S(x)\eqno(12)$$
$$ \tilde S(x)= -{\partial \tilde F\over \partial x}
\eqno(13)$$
The asymptotes of $\tilde S(x) $ at $x\gg 1$ and $x\ll 1$
are
$$ \tilde S(x) \sim \ln x + x^2 ~~,~~x\gg 1\eqno(14)$$
$$ \tilde S(x) \sim  x   ~~,~~x\ll 1\eqno(15)$$
The vortex part of the normalized  entropy $\tilde S(x)$,
i.e.,  without the bulk term $x^2$ in Eq.(14),
can be written using the interpolating formula
$$ \tilde S(x) \sim \ln (x + 1) \eqno(16)$$
which gives both the logarithmic term in Eq.(14) at large
$x$ and linear term in Eq.(15) at low $x$.

It is instructive to write the vortex entropy per vortex
per layer:
$${S_V(T,B) \over k_B}\sim {E_F\over  T_c } \ln (x + 1),
\eqno(17)$$
where $E_F$ is the Fermi energy. Note that the logarithmic
vortex entropy in Eq.(14)  also follows from the $1/E$
behavior of the vortex DOS found in Ref.\cite{KopninVolovik}.
Using the result of Ref.\cite{KopninVolovik} one can find
an exact equation for the vortex entropy at large $x$ using
an axial distribution of the superfluid velocity around the
vortex, ${\bf v}_s= (\hbar/2m_3 r ) \hat\phi$:
$${S_V(T,B) \over k_B}=2\ln 2 ~{v_Fp_F\over  \Delta' } \ln
x ~~,~~x\gg 1 . \eqno(18)$$
Here $\Delta'$ is the angle derivative of the gap at the
node.

\subsection{Heat capacity.}

For the heat capacity one has
$$C(T,B) =T{\partial S\over \partial T}=N_F   T_c {B\over
B_{c2}} \tilde C(x) \eqno(19)$$
$$  \tilde C(x)=x{\partial \tilde S\over \partial
x}\eqno(20)$$
Using the interpolating formula (17) for the entropy, one
obtains interpolating formulas for the vortex part of
the  heat capacity:
$$C(T,B) \sim N_F   T_c {B\over B_{c2}} {x\over 1+x}
,\eqno(21)$$
which gives both asymptotes found in
Ref.\cite{KopninVolovik}:
$\tilde C(x)\sim x$ for $x \ll 1$ and $  \tilde C(x)\sim 1$
for $x \gg  1$.

\section{Discussion.}

The electronic entropy per vortex per layer $S_V$ in
Eq.(17) is much larger than $k_B$ even at $T\ll
T_c$. For an isolated vortex this entropy diverges as the
logarithm of the dimension $R$ of the system: $S_V\sim
k_B(E_F/T_c) \ln R$ (actually $R$ is limited by the
penetration length)  and is at least a factor of $E_F/T_c\gg
1$ larger than the configurational entropy of the vortex in
the 2-dimensional system, $S_{\rm conf}\sim k_B  \ln R$.
The logarithmic behavior of $S_V$, with $R$ limited by the
intervortex distance $R_V$, persists till  $T\sim
T_c\sqrt{B/B_{c2}}$ (or $x \sim  1$). Due to the large
factor $E_F/T_c\gg 1$  the entropy per  vortex per
layer can be of order $k_B$ even at $T<  T_c\sqrt{B/B_{c2}}$
(or $x <  1$), but it finally disappears in the high-field
limit ($T \ll  T_c\sqrt{B/B_{c2}}$ or $x \ll 1$).

It is important that $S_V$ depends on the distribution
of the velocity field ${\bf v}_s$ around the vortex. If
there is a first-order transition with the change of the
velocity
distribution, one can expect a big entropy jump, $\Delta S
\sim 1 k_B$. This entropy jump comes from the electronic
degrees of freedom in the vortex background, which is
modified by the vortex transition. This can explain the
latent heat  $L \sim 0.45  k_BT$/vortex/layer ovserved on
the so called vortex-melting line in a detwinned Y-123 crystal
\cite{Schilling} and $L \sim 0.6 \pm 0.1  k_BT$/vortex/layer
in a twinned sample of Y-123 \cite{Junod}. Even higher values
of the entropy jump have been deduced from the magnetization
measurements \cite{MagnetizationJump}. Such an entropy jump
can  occur both at the vortex-melting transition and at a
first-order transition in which the structure of the vortex
lattice changes, say, from a hexagonal lattice close to
$T_c$ to a distorted tetragonal lattice far from $T_c$.
The latter structure was observed in
Ref.\cite{Maggio-Aprile} and discussed in Ref.\cite{Affleck}.

 Note that the fermionic entropy of the 3-dimensional vortex
loop of the length $R$ is $\propto R \ln R$, as distinct
from the configurational entropy of the loop $S_{\rm conf}
\propto R$. This $ R \ln R$ behavior of the vortex loop
entropy, together with the large prefactor, can in principle
cause the Berezinskii-Kosterlitz-Thouless transition
in the 3-dimensional system.
This is supported by the following observation. It appears
that if one uses the quantum-mechanical approach to the
vortex DOS, by calculating the discrete bound states of the
fermions on the background of the inhomogeneous distribution
of the superflow around the vortex, one obtains a value of
the vortex DOS that is twice as large as that obtained from a
classical treatment of the fermions in terms  of the
Doppler shifted  energy $E + {\bf v}_s({\bf r})\cdot  {\bf p}$
(Ref.\cite{KopninVolovik}).

The classical approach gives the conventional expression for
the  energy of isolated vortex in terms of the superfluid
density $\rho_s(T)=\rho -\rho_n(T)$
$$F_V=  {\pi \hbar^2\over 4m_e^2} (\rho -\rho_n(T))  \ln {R
\over \xi}
~~.\eqno(22)$$
At low $T$, where $\rho_n(T)/\rho\sim T/T_c$,  the second
term corresponds to a logarithmic entropy of the vortex.
The quantum-mechanical approach in terms of the fermionic
bound states in the vortex background suggests the larger
contribution of the fermions to the vortex entropy. This can
be written using the enhancement factor $1+\alpha(T)$
$$F_V=  {\pi \hbar^2\over 4m_e^2} [\rho
-(1+\alpha(T))\rho_n(T)]  \ln {R  \over \xi}
~~.\eqno(23)$$
According to \cite{KopninVolovik} one has $\alpha(0)=1$. If
this value of $\alpha$ persists to higher temperatures, the
energy of an isolated vortex becomes zero at some
temperature $T_V<T_c$, where
$\rho_n(T_V)={1\over 2}\rho$. Thus
at $T_V$ one would have the Berezinskii-Kosterlitz-Thouless
transition in the 3- dimensional system, occuring due to the
essential  contribution of the fermionic bound states to the
vortex  entropy. However, it is more natural to expect that
$\alpha(T)$  decreases continuously  with temperature,
approaching zero value at $T_c$, since the effect of the
bound states should be negligible in the Ginzburg-Landau
region. So, the possibility of the Berezinskii-Kosterlitz-
Thouless transition in this 3D system depends on the details
of the behavior of $\alpha(T)$.

The author acknowledges the Centre de Recherches sur les
tres Basses Temperatures (CNRS, Grenoble) and the
Departement de Physique de la Matiere Condensee (Universite
de Geneve) for hospitality. I would like to thank O.
Fischer, A. Junod, N.B. Kopnin, B. Revaz and M. Dodgson for
stimulating discussions.

\end{document}